\documentclass[review]{elsarticle}

\usepackage{lineno,hyperref}
\modulolinenumbers[5]

\journal{Physica A}

\begin{document}

\begin{frontmatter}

\title{Rate dependence of current and fluctuations in jump models with negative differential mobility}

\author[address1]{Gianluca Teza}
\author[address1]{Stefano Iubini}
\author[address1,address2]{Marco Baiesi}
\author[address1,address2]{Attilio L. Stella}
\author[address3,address4]{Carlo Vanderzande}
\address[address1]{Department of Physics and Astronomy, University of Padova, Via Marzolo 8, I-35131 Padova, Italy}
\address[address2]{INFN, Sezione di Padova, Via Marzolo 8, I-35131 Padova, Italy}
\address[address3]{Faculty of Sciences, Hasselt University, Agoralaan 1 D, 3590 Diepenbeek, Belgium}
\address[address4]{Institute for Theoretical Physics, KULeuven, Celestijnenlaan 200D, 3001 Leuven, Belgium}

\begin{abstract}
Negative differential mobility is the phenomenon in which the velocity of a particle decreases when the force driving it increases. We study this phenomenon in Markov jump models where a particle moves in the presence of walls that act as traps. We consider transition rates that obey local detailed balance but differ in normalisation, the inclusion of a rate to cross a wall and a load factor. We illustrate the full counting statistics for different choices of the jumping rates. We also show examples of thermodynamic uncertainty relations. The variety of behaviours we encounter highlights that negative differential mobility depends crucially on the chosen rates and points out the necessity that such choices should be based on proper coarse-graining studies of a more microscopic description.  \end{abstract}


\end{frontmatter}


\section{Introduction} 
Since there exists at this moment no general theory for systems out of equilibrium, it is usual to study particular nonequilibrium phenomena using specifically designed Markov jump models. The dynamics of these processes follows once the rates are defined. To do this one can use a general principle such as local detailed balance but this is in general not enough to fix the rates. 

For example, the biased random walk is a popular model for describing one-dimensional transport. Local detailed balance fixes the ratio of the rates for the particle to hop to the right, $p$, and left, $q$, as $p/q=e^F$ (where we put $k_BT$ and the distance between neighbouring sites equal to one). Here $F$ is a force that, for example, is produced by an electric field acting on the particle when it is charged or is the force exerted by an optical tweezer when it represents a molecular motor. In the latter case, it has been found that it is necessary to introduce a load distribution factor $\theta$ in the jump rates, $p \sim e^{\theta F}$ and $q \sim e^{(\theta-1)F}$,  in order to get agreement with experiments \cite{kolomeisky2007molecular,lau2007nonequilibrium,kolomeisky2013motor}. This load factor takes into account the change in the microscopic free energy landscape of the motor due to the force. The necessity of this factor cannot be obtained from local detailed balance alone but can be deduced by taking into account information on a more microscopic scale.

This example shows that an investigation of how a microscopic description can lead under coarse graining to a jump process description is in principle needed to assess the precise choice of transition rates. However, without undertaking such an ambitious program, it is important to explore the physical implications that different choices of rates can have by analysing specific prototype models. Such analysis reveals particularly useful when focusing on non equilibrium phenomena whose occurence sensibly depends on these choices.

Here we perform such an analysis in models for negative differential mobility (NDM). This is the phenomenon where increasing a force leads to a decrease of mobility due to, for example, trapping or crowding.  It has been observed in several experiments on electronic properties of materials \cite{aladashvili1988negative,cen2009oxide,li2017strong,nguyen2013bandgap}. It can also occur in gel electrophoresis of polymers \cite{aakerman2002electrophoretic,michieletto2015rings,michieletto2015topological,iubini2018topological} and chemical reaction networks \cite{falasco2018negative}. Negative differential mobility is often studied theoretically in driven lattice gas models \cite{zia2002getting,cleuren2003brownian,ghosh2014giant,leitmann2013nonlinear,basu2014mobility,benichou2014microscopic,baiesi2015role,illien2018nonequilibrium}. The presence of NDM in these models depends crucially on the chosen rates \cite{baiesi2015role}. If the rates in the direction perpendicular to the force are decreasing with $F$, NDM can be observed. When these rates are however constant, as also allowed by local detailed balance, no NDM is present.

In the present paper we introduce a few exactly solvable one-particle Markov models which differ in their transition rates and study how these different rates influence the presence or absence of NDM. The choice of the rates is motivated by qualitative considerations of the way in which different free energy landscapes could influence the coarse grained description. Various landscapes lead to different rates in the coarse grained Markov model. 

We also go beyond the average current and study higher (scaled) cumulants and the whole probability distribution of the current using large deviation theory \cite{touchette2009large}. In this way our work is also a first step in extending the large deviation theory for currents in exclusion processes \cite{derrida2007non,prolhac2010tree,gorissen2012exact} to lattice gases showing NDM. 

This paper is organised as follows. In section 2, we introduce our Markov jump models. In section 3, we review the large deviation approach to current fluctuations. In section 4, we apply this theory and discuss the difference between the models with and without NDM at the level of current distributions. Finally, in section V we present our conclusions.

\section{Models}
We begin by introducing three related Markov jump models in which a particle can hop between the sites of a lattice with two lanes, see Fig. 1. 
The jump models are considered to be discretisations of an underlying microscopic model  in which a particle performs Brownian motion in an energy landscape $V(x,y)- Fx$ where $F$ is the force that puts the system out of equilibrium. 

To be more specific, in Fig. 1, left-hand side, we show contour plots of two energy landscapes that on a coarse grained scale correspond to the jump models shown in the right-hand side. 
High barriers in the energy landscape are considered as walls in the discretised version. In the potential $V_1$ (upper figure) there are only barriers and walls that are oriented perpendicular to the force. For $V_2$  (lower figure) the landscape is more complex and the walls have the shape of a letter T. 

\begin{figure}
\begin{centering}
\includegraphics[width=12cm]{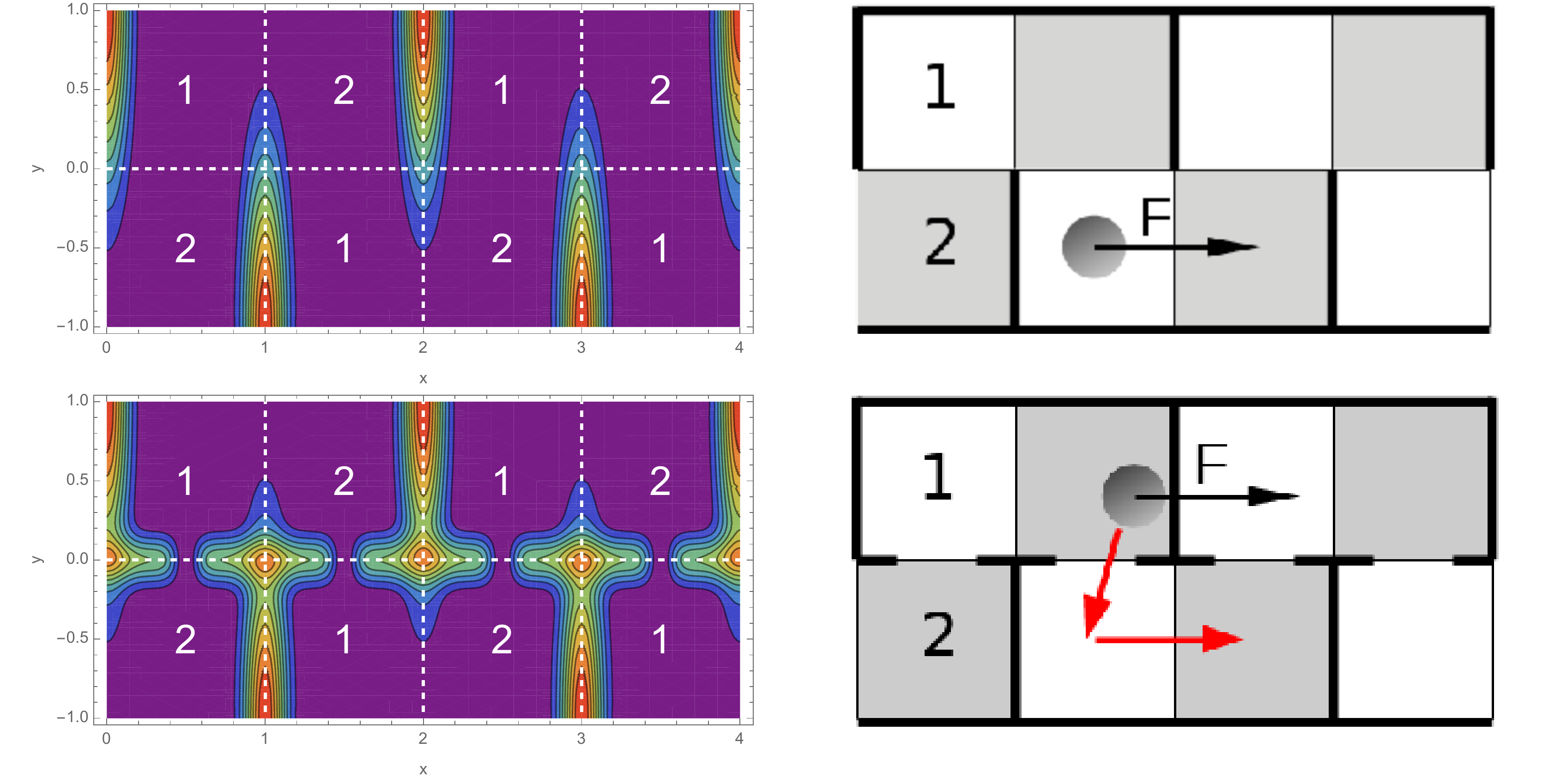}
\caption{The left-hand side shows contour plots of two energy landscapes $V_1$ (top) and $V_2$ (bottom). These are coarse grained to the discretised version on the right-hand side where tick lines are walls that correspond to high energy barriers.}
\end{centering}
\label{fig1}
\end{figure}

The different Markov models are distinguished by different jump rates between neighbouring sites. We denote by $k^+$ and $k^-$ the rates to move in the direction of the field and against the field. As soon as $F \neq 0$, cells with a wall on the right are not equivalent with those where the wall is on the left. Therefore, the jump rates in the direction perpendicular to the force can be different if one goes from a cell with the wall on the left to one with a wall on the right or vice versa. We call the associated rates $u^+$ and $u^-$ respectively. 

We will from now assume that local detailed balance holds even though it is nontrivial to establish a precise connection between its form 
at a microscopic, e.g. Langevin description, and that in the associated jump rate model.

The rates to jump between sites $i$ and $j$ then obey
\begin{eqnarray}
\frac{k(i \to j)}{k(j \to i)}=\exp [E_i - E_j + F \Delta x]
\end{eqnarray}
where $E_i$ is the energy when the particle is at site $i$ and $\Delta x$ is the size of the $x$-component of the jump (which can be $-1,0$ or $1$). Here we will assume that all $E_i$ are equal. This can be the case if we identify the position of the particle in the discrete model with the location of the minima in the potential $V(x)$.
Local detailed balance then implies $k^+/k^-=e^F$ and $k^0\equiv u^+=u^-$. 
 
Given local detailed balance, there is still a large freedom in the choice of the rates. The rate $k^0$ can be $F$-dependent or not. When adding a positive (negative) force, a particle in a cell left (right) of a wall is trapped and first has to change lane in order to move forward. If there is no wall between the two lanes, we do not expect the corresponding rates to depend significantly on $F$ (Fig. 1, top). In the opposite case (Fig. 1, bottom), the particle in the underlying model first has to move against the force to cross the barrier so that we define the corresponding rate $k^0$ to be equal to $1/\cosh{(F/2)}$.

Furthermore we distinguish between models where walls can be passed or not. We introduce an extra (small) parameter $\epsilon$ to describe this effect such that the rates for crossing a wall are given by $k_W^{\pm}=\epsilon k^{\pm}$.  This parameter could be given in terms of a Kramers' rate $\epsilon \sim e^{-\Delta E}$ where $\Delta E$ is the height of the barrier. 

Finally, in order to describe the effect of wall crossing in even more detail, we also allow the possibility of a load-distribution number $\theta$ ($\theta-1$) multiplying $F$ in $k^+$ ($k^-$). As mentioned in the introduction, this takes into account the shift in the location and the height of the energy barriers in the presence of a force. These shifts lead to a modification in Kramers' rate for passing an energy barrier which is captured by the load factor.

With these considerations in mind we define three models. In model A, $k^0$ does not depend on $F$ whereas in the other two it does. In model $B$, $k^0=1/ \cosh{(F/2)}$. In these two models, we do not include a load factor, implying $\theta=1/2$. Finally, in model C, the load factor is an extra parameter.  In Table I, we give a summary of the rates for the various models.
\begin{table}
\begin{center}
\begin{tabular}{l c c c c c }
\hline
\hline
Model & $k^+$ & $k^-$ & $k^0$ \\
\hline
A & $e^{F/2}$ & $e^{-F/2}$ & $1$\\
B & $e^{F/2}$ & $e^{-F/2}$ & $1/\cosh{(F/2)}$ \\
C & $e^{\theta F}$ & $e^{(\theta-1)F}$ & $1/\cosh{(F/2)}$\\
\hline
\end{tabular}
\end{center}
\caption{Jump rates of the particle in the three models.}
\label{Table1}
\end{table}

With the introduction of  periodic boundaries and using the symmetries of the lattice all our models are two state Markov chains. We denote state 1 (2) as the state in which  the particle is to the right  (left) of a vertical wall. The probability to be in state $i$ at time $t$, $P_i(t)$, evolves according to the master equation 
\begin{eqnarray}
\frac{dP_i}{dt}= \sum_{j=1,2} M_{ij} P_j(t)
\label{master}
\end{eqnarray}
A straightforward calculation shows that the generator $M$ is given by
\begin{eqnarray}
M=\left( \begin{array}{cc} -k^0-k^+-k^-_W & k^0+k^-+k^+_W \\ 
k^0+k^++k^-_W & -k^0-k^- -k^+_W\ \end{array}\right)
\end{eqnarray}
The stationary state $P^\star$ is the eigenvector with eigenvalue 0 of $M$. The average current in the stationary state is given by
\begin{eqnarray}
J(F) = (k^+-k_W^-) P^\star(1) -(k^--k^+_W) P^\star(2)
\label{avc}
\end{eqnarray}

We mention here that another popular choice has been to take model $A$ and normalise all rates by a factor $2(1+\cosh{F})$ \cite{zia2002getting,leitmann2013nonlinear,benichou2014microscopic}. From the master equation (\ref{master}), it follows that this can be seen as a rescaling of time to a new force dependent time $t'(F)=t/[2(1+\cosh{F})]$. Since for $F$ large, the rates in this model become constant or go to zero, the current at most goes to a constant. Therefore within such a model it is not possible to obtain current-force relations which  increase with $F$  for $F$ sufficiently large.

We will also be interested in the entropy production rate $\sigma(F)$. A general expression for this quantity for a Markov chain is well known \cite{schnakenberg1976network}. In the present case it reduces to
\begin{eqnarray}
\sigma (F) &=& [(k^+ \ln(k^+/k^-) + k_W^- \ln(k_W^-/k_W^+)] P^\star(1) \nonumber \\ &+&  [(k^- \ln(k^-/k^+) + k_W^+ \ln(k_W^+/k_W^-)] P^\star(2)\end{eqnarray}
Using detailed balance and (\ref{avc}) this can be simplified to
\begin{eqnarray}
\sigma(F) = J(F)  F
\label{ep}
\end{eqnarray}
as could be expected from non equilibrium thermodynamics.

\section{The current large deviation function}
In recent years, much interest has been devoted to go beyond the average current and investigate current fluctuations in various non-equilibrium systems such as exclusion processes \cite{derrida2007non,prolhac2010tree,gorissen2012exact}. Fluctuations of the current away from its average value are characterised in terms of a large deviation function $I(Y)$ which plays a role analogous to that of entropy in equilibrium statistical mechanics. Equivalently, one can describe these fluctuations in terms of the scaled cumulant generating function (SCGF) which is like a non equilibrium free energy. 

Here we collect a few basic results of the theory of large deviations in continuous time Markov processes. For more details we refer to \cite{derrida2007non,lebowitz1999gallavotti}.

In the large deviation approach to current fluctuations, one introduces a stochastic variable $Q(t)$ which increases (decreases) by one each time the particle makes a step in the direction of (against) the field. For $t$ large, the probability density that $Q(t)/t$ equals $Y$ is then proportional to $e^{-t I(Y)}$. Here $I(Y)$ is the large deviation function (LDF, also called rate function) which is zero at the average current $J(F)$ and positive otherwise. 

Alternatively, one can introduce the SCGF as 
\begin{eqnarray}
\lambda(s) = \lim_{t \to \infty} \frac{1}{t} \ln \langle e^{s Q(t)}\rangle
\end{eqnarray}
where the average is taken over all realisations of the process. 
The large deviation function and the SCGF are connected through a Legendre-Fenchel transformation
\begin{eqnarray}
I(Y) = \sup_s [ Y s - \lambda(s)]
\label{legendre}
\end{eqnarray}

The scaled cumulants of the current are obtained by taking derivatives of $\lambda(s)$ with respect to $s$. One has for the first three cumulants 
\begin{eqnarray}
J(F) &=&  \lim_{t \to \infty}\frac{1}{t}\langle Q(t)\rangle= \frac{\partial \lambda(s)}{\partial s} (s=0) \label{av} \\
\Delta (F) &\equiv&  \lim_{t \to \infty}\frac{1}{t}[\langle Q(t)^2\rangle - \langle Q(t) \rangle^2]=\frac{\partial^2 \lambda(s)}{\partial s^2} (s=0) \label{var} \\
\chi (F) &\equiv&  \lim_{t \to \infty}\frac{1}{t}[\langle (Q(t) - \langle Q(t) \rangle)^3\rangle]=\frac{\partial^3 \lambda(s)}{\partial s^3} (s=0) \label{skew}
\end{eqnarray}
For convenience we will refer to the second and third scaled cumulant as the variance and the asymmetry, even though the standard definition of the variance includes an extra factor $t$. 
 
It is known \cite{lebowitz1999gallavotti} that the SCGF equals the largest eigenvalue of a tilted generator $M(s)$
which for the models considered in this paper has the form
\begin{eqnarray}
M(s)=\left( \begin{array}{cc} -k^0-k^+-k^-_W & k^0+k^- e^{-s}+k^+_W e^s  \\ 
k^0+k^+ e^s+k^-_W e^{-s} & -k^0-k^- -k^+_W\ \end{array}\right)
\end{eqnarray}
Here off-diagonal elements that correspond with a step in the direction of the force are multiplied by $e^s$ while those associated to a step against the force get a factor $e^{-s}$.

We have calculated the SCGF for our three models from an exact diagonalisation of the tilted generator. The average, the variance and the asymmetry of the current can then be obtained by simple derivation. 

The average current  (\ref{av}) and the variance (\ref{var}) appear in the thermodynamic uncertainty relation (TUR) \cite{barato2015thermodynamic,proesmans2017discrete,gingrich2016dissipation}
\begin{eqnarray}
\frac{J(F)^2}{\Delta(F)} \leq \frac{\sigma}{2 k_B}
\label{TUR}
\end{eqnarray}
This relation implies that in order to decrease the variance of the current, more entropy has to be produced. 

More recently, also a so called kinetic uncertainty relation (KUR) was derived \cite{di2018kinetic}. It is expressed in terms of the dynamical activity $K(t)$ which, for a given realisation of a jump process, equals the number of transitions made up to time $t$. This dynamic activity, also called frenesy, takes into account non-dissipative aspects of non equilibrium systems and is needed to characterise the physics of systems far from equilibrium \cite{maes2018non,baiesi2009fluctuations}.  The kinetic uncertainty relation is

\begin{eqnarray}
\frac{J(F)^2}{\Delta(F)} \leq \kappa(F)
\label{KUR}
\end{eqnarray}
where $\kappa(F)=\lim_{t \to \infty} \langle K(t) \rangle /t$, i.e. the average activity per unit time.
The KUR might give a better bound on current fluctuations when the system is far from equilibrium \cite{di2018kinetic}, though in a chemical reaction network the reverse situation has been observed \cite{falasco2018negative}. 

In our  models we have
\begin{eqnarray}
\kappa(F) = (k^+ + k^0+ k_W^-) P^\star(1) + (k^- + k^0 + k^+_W) P^\star(2)
\end{eqnarray}

Finally, we have also determined the rate function $I(Y)$ using (\ref{legendre}). For given $Y$ we have to determine $s^\star(Y)$ which satisfies
\begin{eqnarray}
Y = \frac{\partial \lambda(s)}{\partial s} (s^\star)
\label{par}
\end{eqnarray}
Then $I(Y)= Y s^\star(Y) - \lambda(s^\star(Y))$. In practice it is more convenient to vary $s$ and make a parametric plot of the curve $\{ \frac{\partial \lambda(s)}{\partial s}(s), \frac{\partial \lambda(s)}{\partial s}(s) s - \lambda(s)\}$ \cite{touchette2009large}.

\section{Results}
In Fig. \ref{fig2} we plot the average current (top), the variance (middle) and the asymmetry (bottom) as a function of $F$ for the three models. For model A and B (left and middle panel) the results are given for different $\epsilon$-values, while for model C (right panel) results are shown for different $\theta$-values and at $\epsilon=0.005$. 

We find that model A does not show NDM whereas the other two models do. There is also NDM in these models in a regime of negative $F$ (not shown). This is because they have an ($F \leftrightarrow -F$, left $\leftrightarrow$ right) symmetry. 
\begin{figure}
\begin{centering}
\includegraphics[width=14cm]{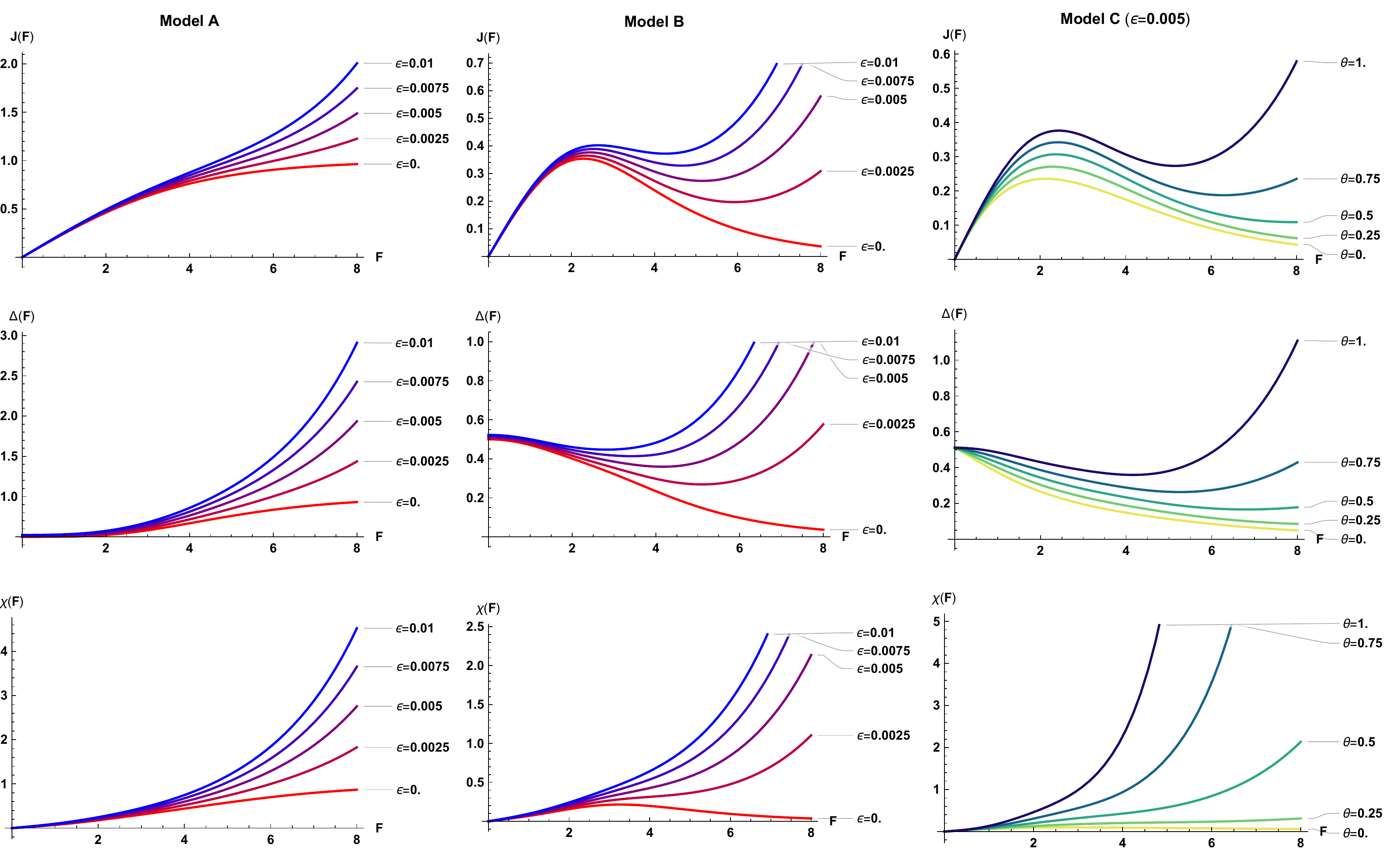}
\end{centering}
\caption{Average current (top), variance (middle) and asymmetry (bottom) as a function of $F$ for various $\epsilon$ values. The left (middle) panel shows the result for model A(B). On the right we show the same quantities for model C at fixed value of $\epsilon=0.005$ as a function of the load factor $\theta$.}
\label{fig2}
\end{figure}

Notice that for model B and for $\epsilon=0$  the average current goes to zero for large $F$ and that NDM is therefore present for all forces above a critical one.  If the walls can be surpassed and $\varepsilon$ is not too big, NDM is present in a finite interval of forces, i.e. for $F_-(\epsilon) \leq F \leq F_+(\epsilon)$ after which the current increases again. For $\epsilon$ above a critical value $\epsilon_c = 0.015$, NDM disappears. In Fig. 3, we present a plot of $F_-(\epsilon)$ and $F_+(\epsilon)$ in model B. 
\begin{figure}
\begin{centering}
\includegraphics[width=8cm]{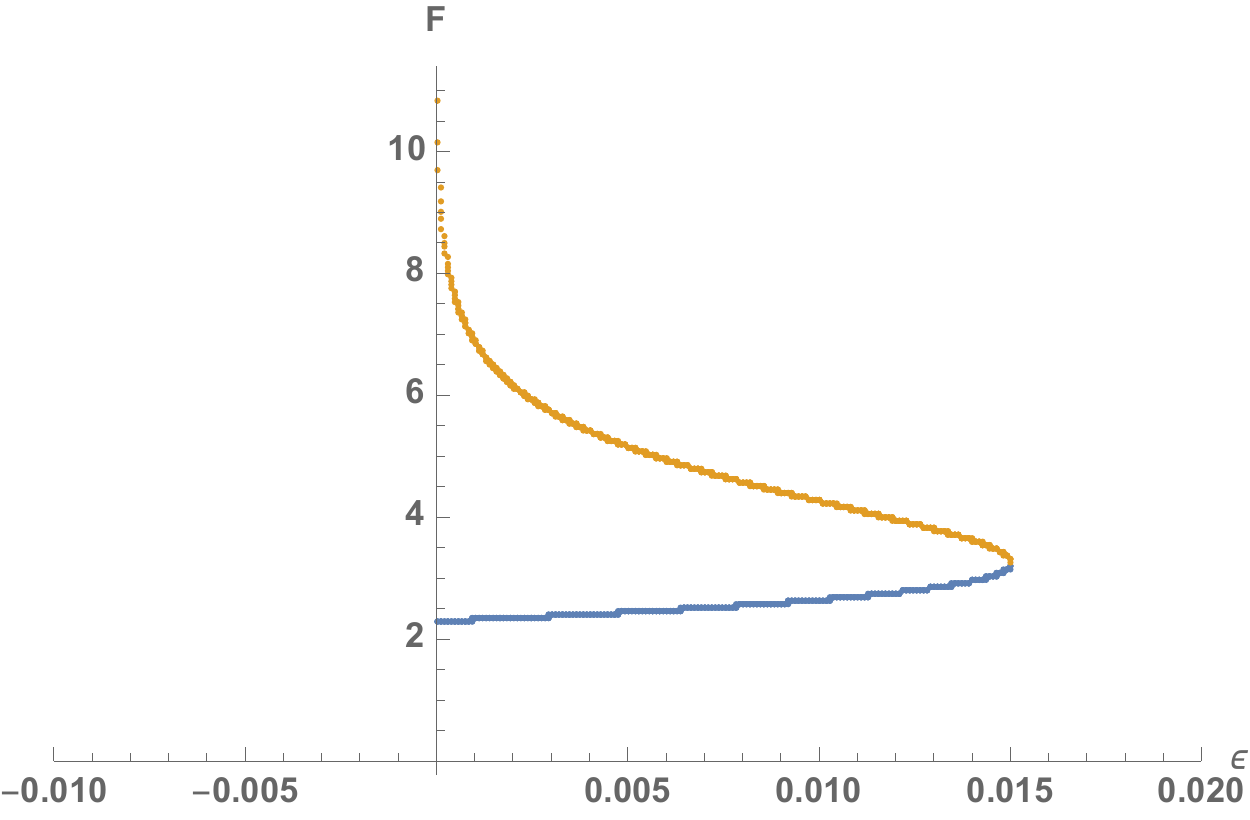}\end{centering}
\caption{This figure shows, for model B, the regime in which there is negative differential mobility. It is bounded by the forces $F_-(\epsilon)$ (blue, lower curve) and $F_+(\epsilon)$ (orange, upper curve). Above $\epsilon =0.0015$, NDM disappears.}
\label{fig2}
\end{figure}

Interestingly, the observed behaviour of the current in model B and C at $\varepsilon \neq 0$ is qualitatively similar to the behaviour seen in some experiments on hopping conductivity \cite{aladashvili1988negative,cen2009oxide,li2017strong,nguyen2013bandgap}. In all of these, it has been found that the current as a function of the field shows NDM for a certain range of forces, after which it increases again. 

If we normalise all the rates of model A with a factor $2(1+\cosh{F})$ \cite{zia2002getting,leitmann2013nonlinear,benichou2014microscopic} one finds that also model A shows NDM. On the other hand, we then find that the current goes to a constant at large force which is not what one would expect physically for a particle in a flow or for the hopping conductivity of an electron. 

We now look whether signs of NDM can also be seen in the fluctuations of the current. In model A, the variance is monotonically increasing both as a function of $F$ and as a function of $\varepsilon$. In contrast, in the models showing NDM, the variance goes to zero at large $F$ for $\varepsilon=0$ while it goes through a minimum if $\varepsilon \neq 0$. Hence in both cases there is a regime in which $\partial \Delta (F)/\partial F <0$. This phenomenon can be called negative differential variance.

Also the asymmetry is observed to behave in a similar way: in the models with NDM there is a regime where $\partial\ \chi(F)/\partial F <0$. Such a regime is absent in model A. Notice however that this behaviour is only observed for $\epsilon$ and $\theta-1/2$ very small. 

\begin{figure}
\begin{centering}
\includegraphics[width=14cm]{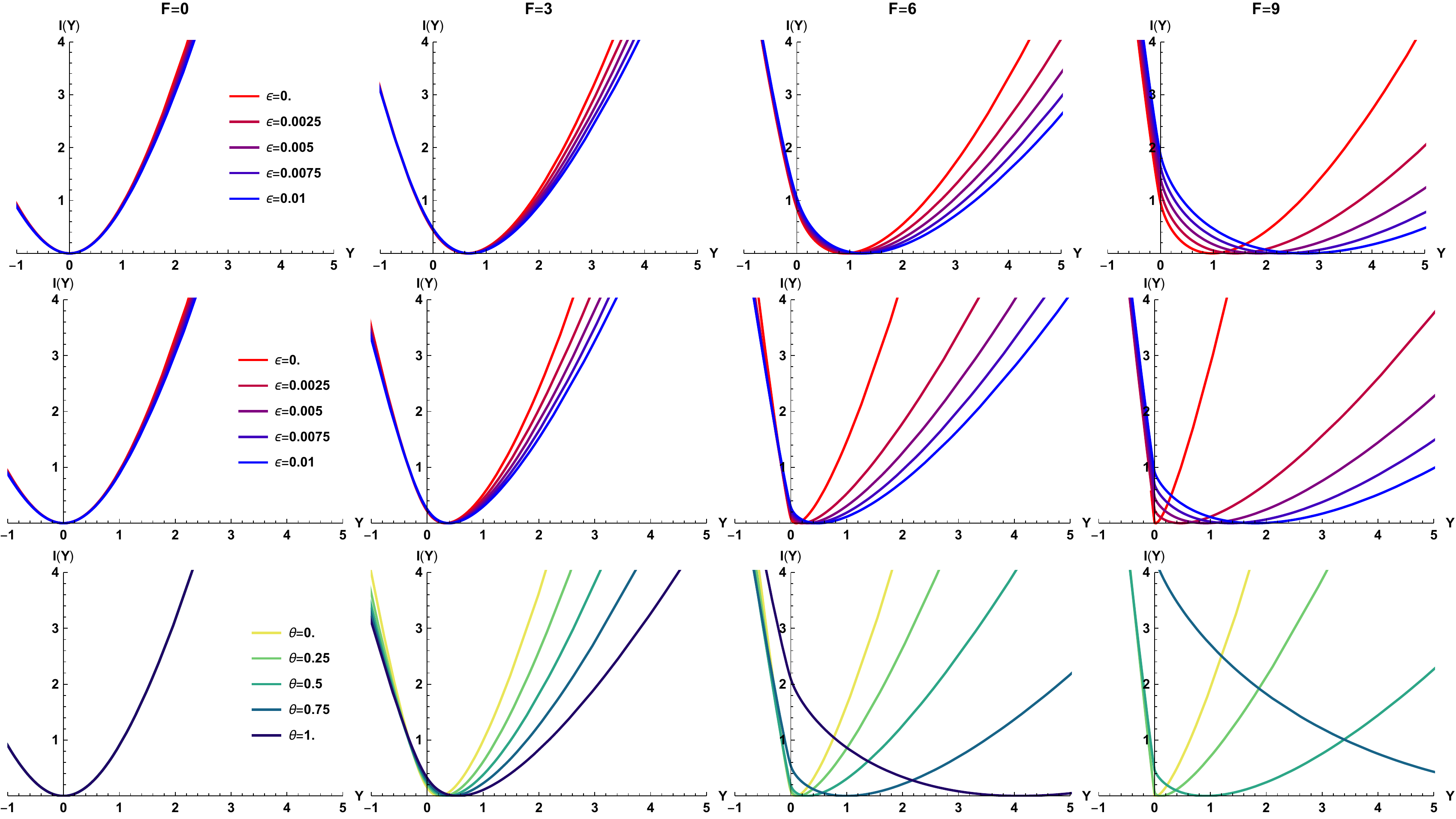}
\end{centering}
\caption{Large deviation function $I(Y)$ in model A (top) and B (middle) at $F=0, 3, 6$ and $9$ for different $\epsilon$-values. The bottom figure shows $I(Y)$ in model C as a function of $\theta$ at $\epsilon=0.005$. }
\label{fig5}
\end{figure}

Alternatively, the same conclusions can be reached by plotting the rate function $I(Y)$. In Fig. \ref{fig5} we plot this function for the three models for various forces and $\epsilon$ or $\theta$ (model C). When the average current increases (decreases) the minimum in $I(Y)$ shifts to the right (left). Similarly, an increase (decrease) in $\Delta (F)$ makes $I(Y)$ broader (smaller) while an increase (decrease) in $\chi (F)$ tilts the function more to the right (left).

\begin{figure}
\begin{centering}
\includegraphics[width=14cm]{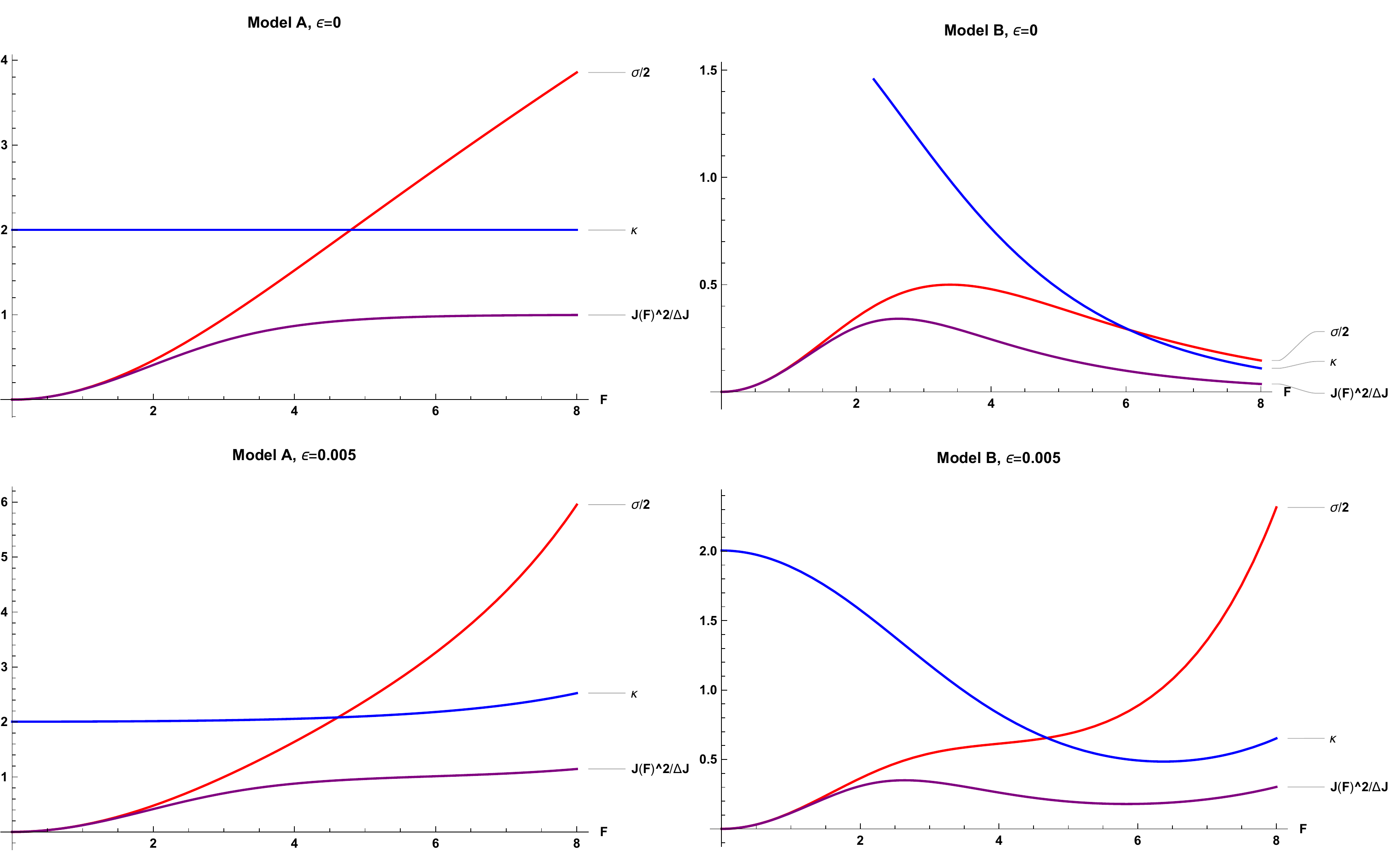}
\end{centering}
\caption{Entropy production rate $\sigma(F)/2$, average activity rate $\kappa(F)$ and $J(F)^2/\Delta J$ as a function of $F$ in model A and model B for $\epsilon=0$ and $\epsilon=0.005$.}
\label{fig6}
\end{figure}

Finally we look at the uncertainty relations (\ref{TUR}) and (\ref{KUR}). In Fig. \ref{fig6} we plot, for models A and B, $\sigma(F)/2$ and $\kappa(F)$ at $\epsilon=0$ and $\epsilon=0.005$. In the same plot we also show $J(F)^2/\Delta J$. We see that the thermodynamic uncertainty relation becomes almost an equality for small forces, while it gives a strict upper bound at large forces. For larger forces, where it can be argued that the system is further from equilibrium, the KUR gives a better upper bound. This result holds independently of the behaviour of the entropy production at large $F$. The same conclusion was found to be valid for the other models and other parameter values. 

\section{Conclusion}
In this paper we have investigated the current fluctuations in some Markov chain models that were constructed to give a coarse grained description of the Brownian motion of a particle in an energy landscape. We have chosen the rates in such a way that they give an average current that behaves similarly to the one expected in the microscopic model. While the different models all obey the same local detailed balance we have seen that choosing different rates can lead to the absence or presence of negative differential mobility.
If we include more parameters in the rates, such as the possibility to cross a wall or a load factor, we can obtain a current-force relation that qualitatively resembles that observed in experimental systems.

Besides average currents we also investigated the current fluctuations. We have seen that in models with NDM, and for $\epsilon=0$ also the variance and the asymmetry of the current have a maximum which for the variance is at $F=0$. When $\varepsilon \neq 0$ the behaviour becomes more interesting so that for some parameter values the variance and the asymmetry go through a minimum as a function of $F$.
This can be compared with lattice gas models of NDM. In these one considers one or more force driven particles in the presence of passively diffusing particles that do not feel any force. The latter can act as traps and cause NDM \cite{benichou2014microscopic,baiesi2015role}. It has been found that for such a model the variance of the current goes through a maximum \cite{illien2018nonequilibrium}. It would be interesting to see whether with a suitable modification of the lattice gas model one could also find regimes in which the average current and its fluctuations show both a maximum and a minimum. 

In this work, the relation between the microscopic and the mesoscopic model was made at a heuristic level. It would be of great interest to make a more detailed mapping between the two levels of description. One can for example ask how the rates should be chosen such that the models at different scales have the same current and the same diffusion constant. Or one can require that the entropy production at the two levels stays the same. These questions have been studied in an approximate way for Markov chain models in which a clear separation of timescales can be made between fast and slow variables \cite{rahav2007fluctuation,puglisi2010entropy}. But that work leaves open the question on what happens if one coarse grains a Markov chain where no clear separation of time scales is present, or when one coarse grains from a microscopic diffusion process to a mesoscopic Markov model. We have recently performed a study of these issues and results will be published elsewhere.

\section*{References}

\bibliography{references}

\end{document}